\documentclass[twocolumn,showpacs,preprintnumbers,amsmath,amssymb,prl]{revtex4-1}
\usepackage{mathrsfs}
\usepackage{graphicx}
\usepackage{dcolumn}
\usepackage{bm}
\usepackage{amsmath}
\usepackage{amsfonts}

\begin{document}

\title{Anomalous Edge Transport in the Quantum Anomalous Hall State}
\author{Jing Wang}
\author{Biao Lian}
\author{Haijun Zhang}
\author{Shou-Cheng Zhang}
\affiliation{Department of Physics, McCullough Building, Stanford University, Stanford, California 94305-4045, USA}

\begin{abstract}
We predict by first-principles calculations that thin films of Cr-doped (Bi,Sb)$_2$Te$_3$ magnetic topological insulator have gapless non-chiral edge states coexisting with the chiral edge state. Such gapless non-chiral states are not immune to backscattering, which would explain dissipative transport in the quantum anomalous Hall (QAH) state observed in this system experimentally. Here we study the edge transport with both chiral and non-chiral states by Landaur-B\"{u}ttiker formalism, and find that the longitudinal resistance is nonzero whereas Hall resistance is quantized to $h/e^2$. In particular, the longitudinal resistance can be greatly reduced by adding an extra floating probe even if it is not used, while the Hall resistance remains at the quantized value. We propose several transport experiments to detect the dissipative non-chiral edge channels. These results will facilitate the realization of pure dissipationless transport of QAH states in magnetic topological insulators.
\end{abstract}

\date{\today}


\maketitle

\paragraph{Introduction}

The recent theoretical prediction and experimental realization~\cite{qi2006,qi2008,liu2008,li2010,yu2010,chang2013b} of QAH effect have generated intense interest in this new state of quantum matter. The QAH insulator has a topologically nontrivial electronic structure characterized by a bulk energy gap but gapless chiral edge states, leading to the quantized Hall effect without an external magnetic field~\cite{haldane1988}. In the quantum Hall effect (QHE), electronic states of two-dimensional (2D) electron system form Landau levels under strong external magnetic field, and the Hall resistance is accurately quantized into $h/\nu e^2$ plateaus~\cite{klitzing1980,tsui1982} accompanied by exact zero longitudinal resistance and conductance in the plateaus (here $h$ is Plank's constant, $e$ is the charge of an electron, and $\nu$ is an integer or a certain fraction). The exact quantization of the Hall resistance arises from dissipationless chiral states localized at sample edges~\cite{laughlin1981}, along which electric currents flow unidirectionally and backscattering cannot take place~\cite{halperin1982}. In a QAH insulator, theoretically predicted in magnetic topological insulators (TIs)~\cite{qi2006,qi2008,liu2008,li2010,yu2010}, the spin-orbit coupling (SOC) and ferromagnetic ordering combine to give rise to a topologically nontrivial phase characterized by a finite Chern number~\cite{thouless1982} and chiral edge states characteristic of the QAH state. The QAH effect has been experimentally observed in thin films of Cr-doped (Bi,Sb)$_2$Te$_3$ magnetic TI~\cite{chang2013b}, where at zero magnetic field, the gate-tuned Hall resistance ($\rho_{xy}$) exhibits a plateau with quantized value $h/e^2$ while the longitudinal resistance ($\rho_{xx}$) shows a dip down to 0.098~$h/e^2$. This quantized value of $\rho_{xy}$ is consistent with quantum transport due to a single chiral edge state. However, nonzero $\rho_{xx}$ indicates that the system has other dissipative conduction channels. Thus, it is important to be able to trace where such dissipation come from and to realize experimentally a pure dissipationless transport of QAH states in magnetic TIs~\cite{wang2013}.

In this paper, based on first-principles calculations, we show that 5 quintuple layers (QLs) of Cr$_x$(Bi,Sb)$_{2-x}$Te$_3$ magnetic TI studied in the experiment~\cite{chang2013b} has gapless non-chiral edge states coexisting with chiral edge state. Such gapless non-chiral states are not immune to backscattering, which would explain dissipative transport in the QAH state observed in magnetic TI~\cite{chang2013b}. Here we study the edge transport with both chiral and non-chiral states by Landaur-B\"{u}ttiker formula, and find that $\rho_{xx}$ exhibits non-ohmic behavior. Remarkably, $\rho_{xx}$ is nonzero whereas $\rho_{xy}$ is quantized into $h/e^2$. In particular, $\rho_{xx}$ can be greatly reduced by the mere presence of a floating probe even if it is not used, while $\rho_{xy}$ remains at the quantized value. The non-chiral edge channels can be detected through nonlocal transport measurements. We also predict that thinner films of Cr-doped (Bi,Sb)$_2$Te$_3$ is a QAH insulator with one single chiral edge state, in which pure dissipationless transport of QAH states can be realized.

\paragraph{Materials}
\begin{figure*}[t]
\begin{center}
\includegraphics[width=5.3in]{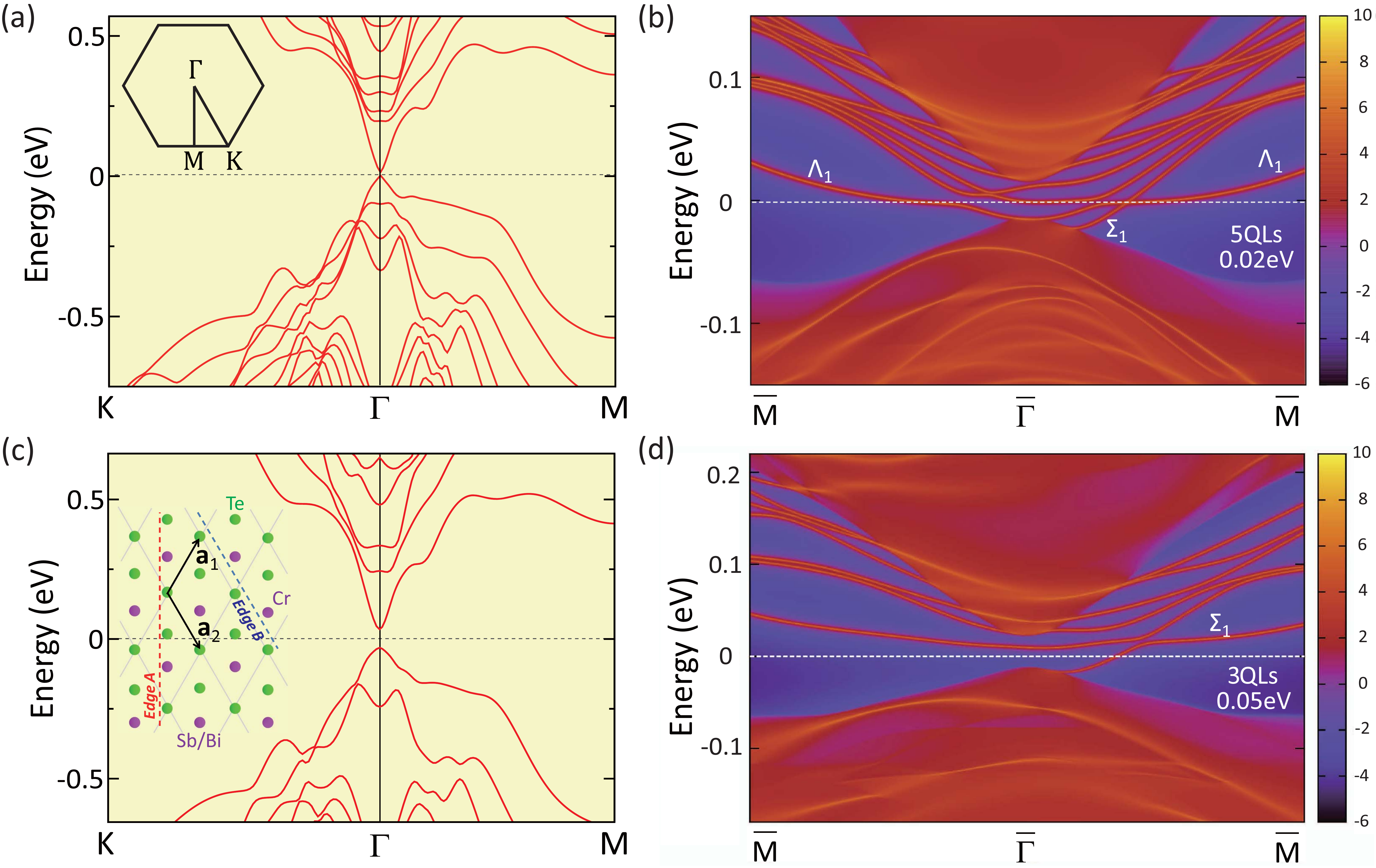}
\end{center}
\caption{(color online) Band structure for 5-QLs and 3-QLs Cr$_{0.15}$(Bi$_{0.1}$Sb$_{0.9}$)$_{1.85}$Te$_{3}$ without exchange field is plotted in (a) and (c), respectively. The dashed line indicates the Fermi level. The inset of (a) shows the 2D Brillouin zone with high-symmetry \textbf{k} points $\Gamma$(0,0), K($\pi$,$\pi$) and M($\pi$,0) labelled and that of (c) is the top view of 2D thin film with two in-plane lattice vectors $\mathbf{a}_1$ and $\mathbf{a}_2$. The 1D edges are indicated by the dashed lines, \emph{Edge A} and \emph{Edge B}. The energy dispersion of the semi-infinite Cr$_{0.15}$(Bi$_{0.1}$Sb$_{0.9}$)$_{1.85}$Te$_{3}$ film along \emph{Edge A} is plotted for (b) 5-QLs with exchange field 0.02~eV, (d) 3-QLs with exchange field 0.05~eV. Here, the warmer colors represent the higher LDOS, with red and blue regions indicating 2D bulk energy bands and energy gaps, respectively. The gapless edge states can be clearly seen around the $\Gamma$ point as red lines dispersing in the 2D bulk gap. One gapless chiral edge state $\Sigma_1$ and one pair of gapless quasi-helical edge states $\Lambda_1$ coexist in (b), while only one gapless chiral edge state $\Sigma_1$ exists in (d).}
\label{fig1}
\end{figure*}

We study the Cr-doped (Bi$_{0.1}$Sb$_{0.9}$)$_2$Te$_3$ magnetic TI, where the Dirac cone of surface states is observed to be located in the bulk band gap~\cite{chang2013b,zhang2011}. Here, we carry out first-principles calculations on three-dimensional (Bi$_{0.1}$Sb$_{0.9}$)$_2$Te$_3$ without SOC. The virtual crystal approximation is employed to simulate the mixing between Bi and Sb in first-principles calculations. Then we construct the tight-binding model with SOC and the exchange interaction based on maximally localized Wannier functions~\cite{marzari1997,souza2001}. The effective SOC parameter of Cr$_{x}$(Bi$_{0.1}$Sb$_{0.9}$)$_{2-x}$ is obtained by linear interpolation between the SOC strength of Bi and Sb, where the reduced SOC strength resulting from the Cr substitution of (Bi, Sb) has been taken into account~\cite{zhang2013}. When the 2D system stays in the QAH phase, there must be chiral edge states if an edge is created. Here we study the edge states of Cr$_x$(Bi,Sb)$_{2-x}$Te$_3$ thin films along \emph{Edge A} direction, as shown in Fig.~\ref{fig1}. For a semi-infinite system, combining the tight-binding model with the iterative method~\cite{sancho1984}, we can calculate the Green's function for the edge states directly. The local density of states (LDOS) is related to the imaginary part of Green's function, from which we can obtain the dispersion of the edge states. As shown in Fig.~\ref{fig1}(b) for 5-QLs Cr$_{0.15}($Bi$_{0.1}$Sb$_{0.9}$)$_{1.85}$Te$_3$, there indeed exists one chiral edge state $\Sigma_1$ indicating the $\nu=1$ QAH state. There are also other trivial edge states, but most of them only connect to the conduction or valence band. Remarkably, one pair of these trivial edge states $\Lambda_1$ is gapless, which connects the conduction and valence band.

$\Lambda_{1}$ can be dubbed as the \emph{quasi-helical} edge states. It originates from helical edge states of quantum spin Hall (QSH) effect but with time-reversal symmetry (TRS) breaking due to spontaneous magnetic moments, where the gap is opened at Dirac point and buried into valence bands by particle-hole asymmetry. It is non-chiral, with two counterpropagating channels, but not immune to backscattering due to TRS breaking. Such coexistence of chiral and quasi-helical edge states is quite general in magnetic TIs, especially in thick films. These quasi-helical states do not change the topological property of the system, however, they contribute to the dissipative edge transport and can be used to explain nonzero $\rho_{xx}$ when $\rho_{xy}$ is quantized in the QAH experiment~\cite{chang2013b}.

\paragraph{Edge transport}

\begin{figure*}[t]
\begin{center}
\includegraphics[width=5.4in]{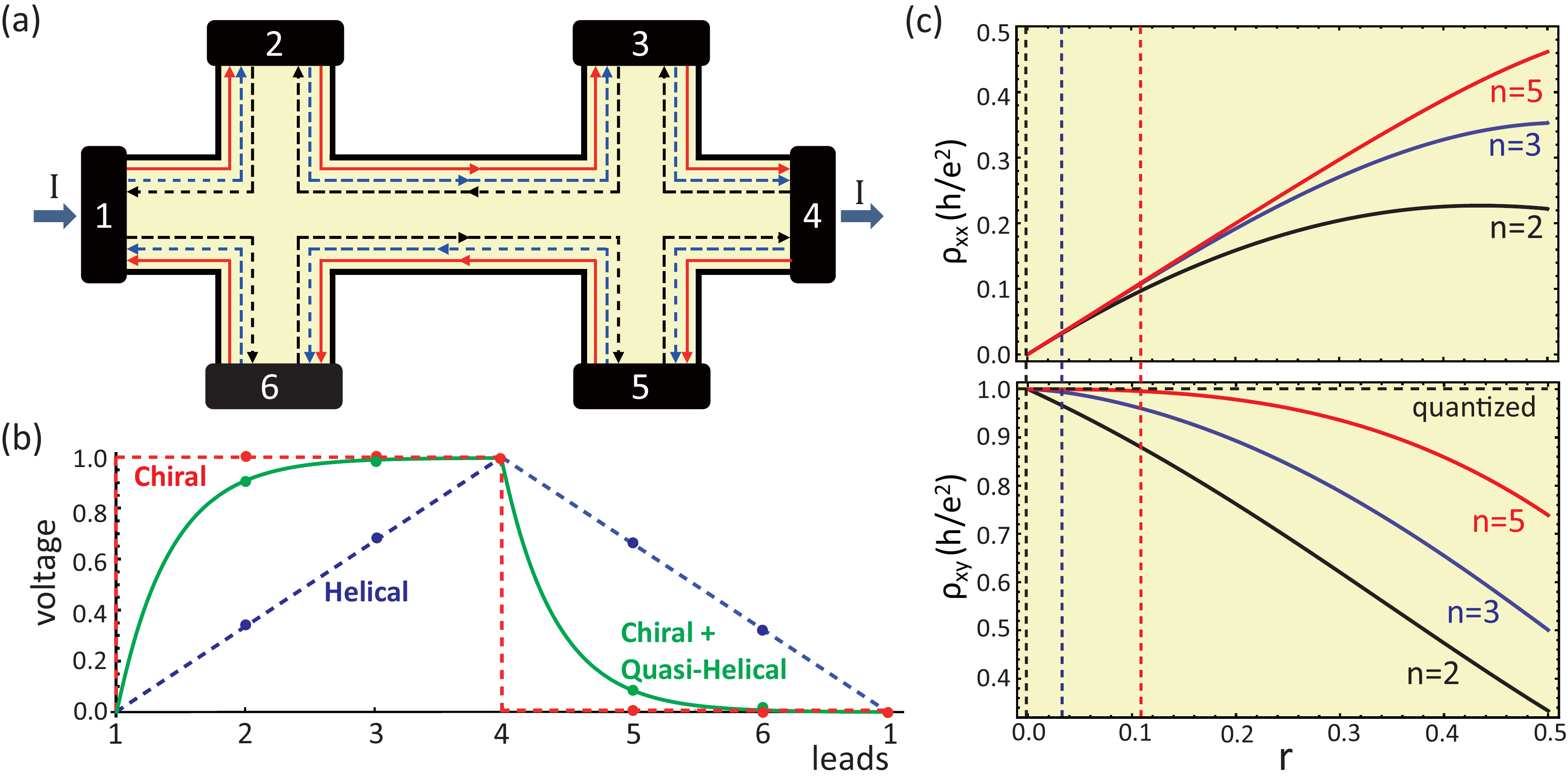}
\end{center}
\caption{(color online) Hall bridge and transport properties. (a) Schematic drawing of a Hall bar device with both quasi-helical edge channels (dashed blue and black) and chiral edge channel (solid red). The current is from terminal 1 to 4, voltage leads are on electrodes 2, 3, 5, and 6. (b) Voltage at terminal 1-6 for transport of chiral, helical, chiral \& quasi-helical edge state. The transport with both chiral and quasi-helical edge channels (solid green) show non-ohmic behaviors of $\rho_{xx}$. (c) $\rho_{xx}$ and $\rho_{xy}$ vs. $r$ with different numbers of effective voltage leads on each side of the sample.}
\label{fig2}
\end{figure*}

To demonstrate the existence of predicted extended non-chiral edge channels in magnetic TI, we study the edge transport with both chiral and non-chiral states by Landaur-B\"{u}ttiker formalism~\cite{buttiker1986,buttiker1988}. The general relationship between current and voltage is expressed as
\begin{equation}\label{LBformula}
I_i=\frac{e^2}{h}\sum\limits_{j}\left(T_{ji}V_i-T_{ij}V_j\right),
\end{equation}
where $V_i$ is the voltage on the $i$th electrode, $I_i$ is the current flowing out of the $i$th electrode into the sample, and $T_{ji}$ is the transmission probability from the $i$th to the $j$th electrode. There is no net current ($I_j=0$) on a voltage lead or floating probe $j$, and the total current is conserved, namely $\sum_iI_i=0$. The current is zero when all the potentials are equal, implying the sum rules $
\sum_iT_{ji}=\sum_iT_{ij}$.

For a standard Hall bar with $\mathcal{N}$ current and voltage leads [such as Fig.~\ref{fig2}(a) with $\mathcal{N}=6$], the transmission matrix elements for the chiral state of the $\nu=1$ QAH effect are given by $T_{i+1,i}=1$, for $i=1,...,\mathcal{N}$, and others $=0$ (Here we identify $i=\mathcal{N}+1$ with $i=1$). For quasi-helical states, $T_{i+1,i}=k_1$, $T_{i,i+1}=k_2$ and others $=0$. These states are not protected from backscattering and the transmission from one electrode to the next is not perfect, implying $k_1, k_2<1$, which is different from helical edge states in QSH effect where $k_1=k_2=1$~\cite{roth2009}. For simplicity, we have assumed $T_{ij}$ to be translational invariant, namely $T_{ij}=T_{i+1,j+1}$. In general, $k_1$ and $k_2$ become zero for infinitely large sample, because dissipation occurs once the phase coherence is destroyed in the metallic leads or the momentum is relaxed when the sample size $L$ $\gg$ the mean free path $l_{\mathrm{m}}$ ($k_1, k_2\sim l_{\mathrm{m}}/(l_{\mathrm{m}}+L)$). By contrast, the QAH chiral edge states are robust against phase decoherence. Thus the nonzero total transmission matrix elements are
\begin{equation}\label{transmission}
T_{i+1,i} = 1+k_1,\ \ T_{i,i+1} = k_2.
\end{equation}

In the case of current leads on electrodes 1 and 4, and voltage leads on electrodes 2, 3, 5, and 6 as shown in Fig.~\ref{fig2}(a), one finds that $I_1=-I_4\equiv I$, and the voltage from 1, 2, 3, to 4 increases exponentially, whereas the voltage from 4, 5, 6, to 1 decreases exponentially,
\begin{eqnarray}
V_j &=& \frac{1-r^{j-1}}{1-r^3}V, \ \ 1 \leq j\leq 4,
\\
V_j &=& \frac{1-r^{j-7}}{1-r^{-3}}V, \ \ 4 \leq j\leq 6.
\end{eqnarray}
Here we set $V_1\equiv 0$ and $V_4\equiv V$, and $r\equiv k_2/\left(1+k_1\right)$. If $k_1=k_2=0$, which is the case for chiral edge state transport in QAH effect and QHE, $V_2=V_3=V=(h/e^2)I$ and $V_5=V_6=0$, so that $\rho_{xy}\equiv\left(V_2-V_6\right)/I=h/e^2$ and $\rho_{xx}\equiv\left(V_3-V_2\right)/I=0$ as expected [shown in Fig.~\ref{fig2}(b)]. For the helical edge state transport in QSH effect with $T_{i+1,i}=T_{i,i+1}=1$, $V_2=V_6=V/3=(h/2e^2)I$ and $V_3=V_5=2V/3$, and thus $R_{14,14}\equiv\left(V_4-V_1\right)/I=3h/2e^2$ and $R_{14,23}\equiv\left(V_3-V_2\right)/I=h/2e^2$~\cite{roth2009}. For the edge transport with both chiral and quasi-helical states, the voltages of different leads is plotted in Fig.~\ref{fig2}(b), where $\rho_{xx}$ does not scale linearly with the spacing between the voltage leads in accordance with Ohm's law. Moreover, $\rho_{xx}$ is nonzero while $\rho_{xy}$ is nearly quantized. This is the key result of this paper. The sample size in experiment is $>$ 200~$\mu$m, which is much larger than phase coherence length $l_{\phi}<1~\mu$m in this material with a rather low mobility ($<800$~cm$^2$/Vs)~\cite{chang2013b,liu2011}. The effect of decoherence between two real leads can be modeled as an extra floating lead, in which quasi-helical states interact with infinitely many low-energy degrees of freedom, completely losing their phase coherence~\cite{roth2009}. Thus the standard Hall bar with $\mathcal{N}=6$ current and voltage leads [shown in Fig.~\ref{fig2}(a)] used in experiment has effectively $n=5$ voltage leads on each side. As shown explicitly in Fig.~\ref{fig2}(c), for certain parameter range of $r$, $\rho_{xy}$ can be quantized to $h/e^2$ plateau whereas $\rho_{xx}$ is nonzero. This explains the dissipative longitudinal transport of QAH effect observed in magnetic TI recently~\cite{chang2013b}. In a strong external magnetic field $B$, when the magnetic length $l_{\text{B}}<l_{\text{m}}$ ($l_\text{B}\sim \sqrt{\hbar/eB}\sim10$~nm at 10~T), both $k_1$ and $k_2$ approach zero, and thus $\rho_{xx}$ vanishes completely~\cite{chang2013b}.

\begin{figure}[t]
\begin{center}
\includegraphics[width=3.2in]{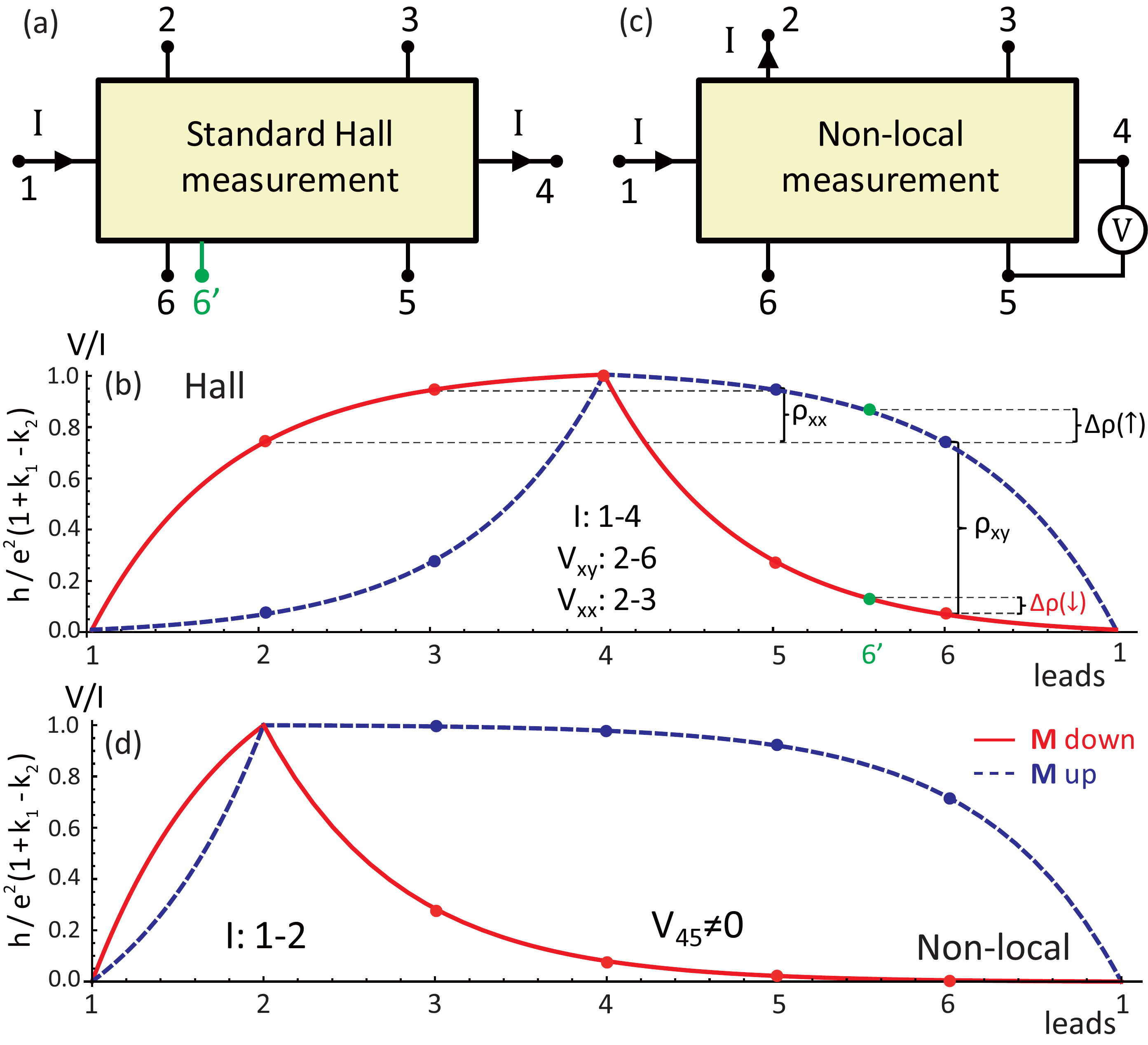}
\end{center}
\caption{(color online) Six-terminal Hall and nonlocal measurements. (a) Standard Hall measurement with six terminals and (b) corresponding voltages.
The current is through 1 to 4, and the Hall voltage is measured between 2 and 6. Terminal $6$ (denoted as $6'$) is not be symmetric to terminal 2 due to misalignment, so Hall signal may contain some longitudinal component. (c) Nonlocal measurement and (d) voltage. The current is through 1 to 2. In (b) and (d), the voltage with downward and upward magnetic orderings are denoted as solid red and dashed blue line, respectively.}
\label{fig3}
\end{figure}

In reality, voltage leads may not be correctly aligned experimentally. As illustrated in Fig.~\ref{fig3}(a), where the current leads are on electrodes 1 and 4. Suppose electrodes 2 and $6'$ are voltage leads in experiment, while position $6$ is the symmetric point (mirror) of 2. The voltage of leads in this Hall bar is plotted in Fig.~\ref{fig3}(b). The solid line and dashed line denotes voltage of leads when magnetization $\mathbf{M}$ is up ($\uparrow$) and down ($\downarrow$), respectively. If the leads are symmetric,
\begin{eqnarray}
\rho_{xy}(\uparrow) &=& \frac{V_2^\uparrow-V_6^\uparrow}{I}=-\frac{V_2^\downarrow-V_6^\downarrow}{I}=-\rho_{xy}(\downarrow)\equiv\rho_0.
\nonumber
\end{eqnarray}
If the leads are not symmetric, namely 6 is moved to $6'$, effectively, such misalignment of leads will cause $V_{6'}$ to be higher than $V_6$ independent of magnetization. So the Hall resistance $\rho_{xy}'$ measured between 2 and $6'$ will gain a fraction of the longitudinal resistance ,
\begin{eqnarray}
\rho_{xy}'(\uparrow) &=& \rho_0-\frac{V_{6'}^\uparrow-V_{6}^\uparrow}{I}=\rho_0-\Delta\rho(\uparrow),
\\
\rho_{xy}'(\downarrow) &=& -\rho_0-\frac{V_{6'}^\downarrow-V_{6}^\downarrow}{I}=-\rho_0-\Delta\rho(\downarrow).
\end{eqnarray}
Thus $\rho_{xy}'(\uparrow)\neq-\rho_{xy}'(\downarrow)$. To the lowest order $\Delta\rho(\uparrow)\approx\Delta\rho(\downarrow)$, and one can anti-symmetrize the Hall resistance to eliminate the effect of asymmetric leads,
\begin{equation}
\rho_{xy} = \frac{\rho_{xy}'(\uparrow)-\rho_{xy}'(\downarrow)}{2}.
\end{equation}
For a well quantized $\rho_{xy}$, one of $\rho_{xy}'(\uparrow)$ and $\rho_{xy}'(\downarrow)$ will be larger than $h/e^2$, while the other will be smaller. This is exactly the phenomena observed in the QAH experiment~\cite{chang2013b}. It worths mentioning that in this system $\Delta\rho(\uparrow)\neq\Delta\rho(\downarrow)$, so this anti-symmetrization process does not cancel the asymmetry effect completely.

\paragraph{Nonlocal transport}

\begin{figure}[tp]
\begin{center}
\includegraphics[width=3.1in]{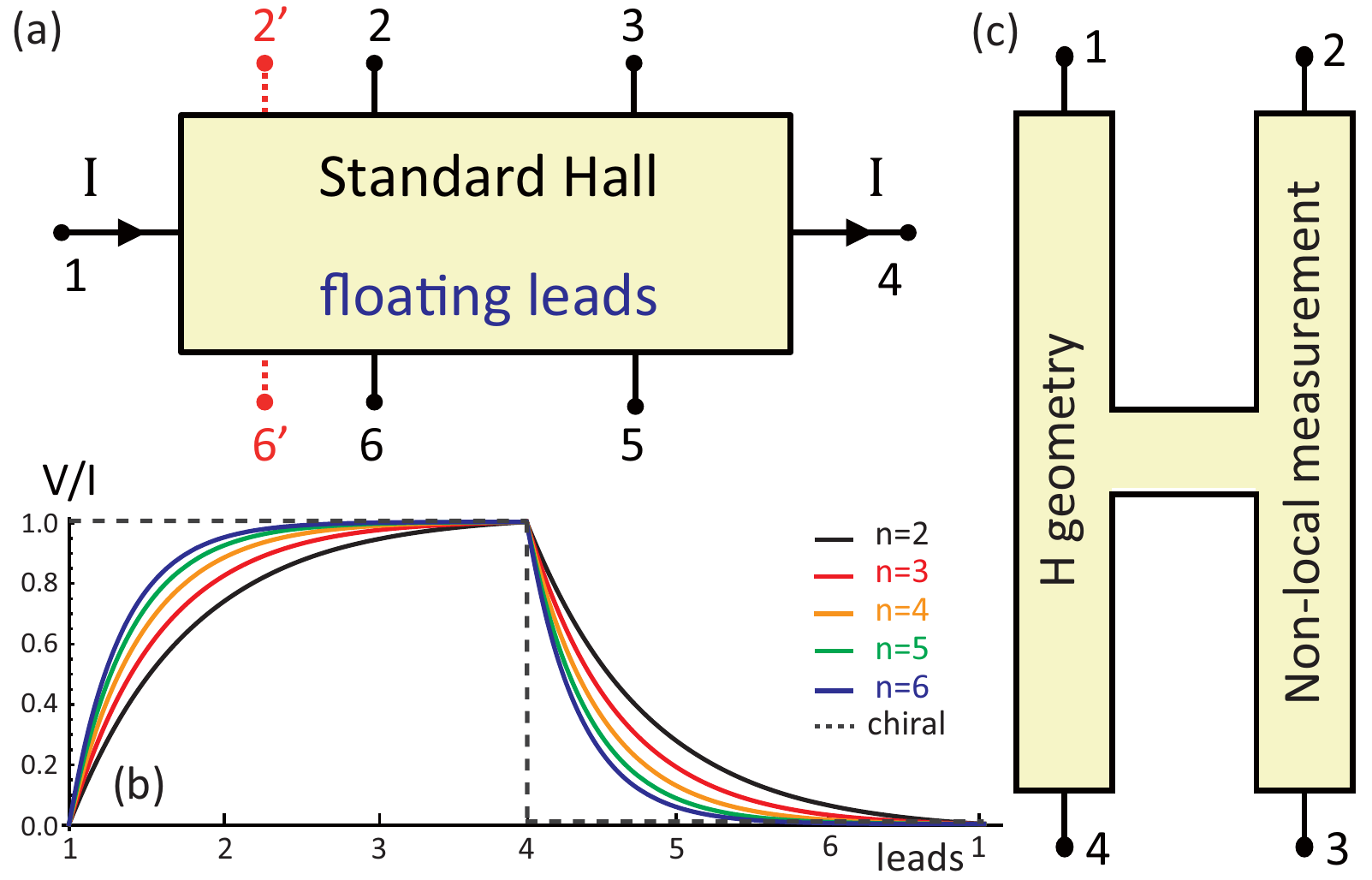}
\end{center}
\caption{(color online) Transport measurements with different numbers of terminals and device geometry. (a) Standard Hall measurement with extra floating terminals $2'$ and $6'$ inserted at edges. $I$: 1-4 and $V_{xy}$: 2-6, $V_{xx}$: 2-3. (b) The voltage at terminals 1-6 of (a) in the presence of extra floating probes. $n$ denotes the total numbers of voltage probes on one side. Dashed line denotes the chiral edge state transport, which is not affected by extra floating probes. (c) Nonlocal four-terminal resistance and two-terminal resistance measurement on the H-bar device.}
\label{fig4}
\end{figure}

The dissipative transport measured in the standard Hall bar does not allow us to distinguish experimentally between quasi-helical edge channels and residual bulk conduction channels in a convincing manner. An unambiguous way to prove the existence of quasi-helical edge
state transport in the QAH experiment is to use nonlocal electrical measurements. The edge states necessarily lead to nonlocal transport, and such nonlocal transport has been experimentally observed in QHE~\cite{buttiker1986,beenakker1991}, which provide definitive evidence for the existence of chiral edge states in QHE.

As shown in Fig.~\ref{fig3}(c) \& (d), a current is passed through probes 1 and 2 and voltage is measured between probes 4 and 5 away from the dissipative bulk current path. For chiral edge state transport, the voltage signal tends to zero. However, for the transport of quasi-helical edge states, $V_4-V_5\neq0$, which gives a nonlocal resistance signal $R_{12,45}/\rho_{xx}\approx0.1$ (around 220~$\Omega$). Here $\rho_{xx}$ is the longitudinal resistance measured by current flowing through 1 to 4 and the voltage between 2 and 3. The classical Ohmic bulk contribution to the nonlocal effect is given by $\mathcal{R}_{\mathrm{NL}}^{\mathrm{classical}}/\rho_{xx}\approx\exp(-\pi\ell/w)$, where $\ell$ is the distance between the voltage probes, and $w$ is the strip width~\cite{pauw1958}. For the geometry with $\ell/w=2$, we estimate $\mathcal{R}_{\mathrm{NL}}^{\mathrm{classical}}/\rho_{xx}\approx10^{-3}$ (5~$\Omega$). Therefore, one would only expect a minimal signal from a conducting bulk. Different from bulk conduction, the quasi-helical edge states are fully nonlocal, and this signature can be taken as a strong evidence for the existence of quasi-helical edge channels transport in the QAH experiment. One can further measure the voltage between electrodes 3 and 4, also 5 and 6. Quantitatively, for edge transport $(V_3-V_4)/(V_4-V_5)=(V_4-V_5)/(V_5-V_6)$, which can further verify the extra dissipative edge channels in magnetic TI. A similar nonlocal voltage can also be studied in a different geometry, for exmaple in the shape of letter H as shown in Fig.~\ref{fig4}(c). The current leads on 1 and 4 and voltage leads on 2 and 3.

Another transport measurement that could directly confirm the existence of quasi-helical edge channels is shown in Fig.~\ref{fig4}(a),
where extra floating probes $2'$ and $6'$ are added to the standard Hall bar~\cite{datta1995}. For the $\nu=1$ QAH effect in magnetic TI, such extra floating leads at sample edges will not affect the transport of residual bulk conduction channels, if there are any. Neither will it affect the chiral edge channel transport. However, it will establish an equilibrium between the two counterpropagating channels of the quasi-helical edge states and changes $\rho_{xx}$ and $\rho_{xy}$. By adding more extra floating probes, $\rho_{xx}$ approaches 0 and $\rho_{xy}$ is more accurately quantized into $h/e^2$, as illustrated in Fig.~\ref{fig4}(b). This is a rather sharp feature which is easy to implement in experiments.

In summary, the coexistence of chiral and quasi-helical edge channels in magnetic TI can explain the dissipative longitudinal transport of the recent QAH experiment. Such quasi-helical edge states can be detected by nonlocal transport measurements. In fact, thinner films of magnetic TIs such as 3-QLs Cr$_{0.15}($Bi$_{0.1}$Sb$_{0.9}$)$_{1.85}$Te$_3$ is a QAH insulator with a single chiral edge state as shown in Fig.~\ref{fig1}(d). There is no gapless trivial edge state in this system, and one can realize the completely dissipationless transport of QAH states.

We acknowledge C. Z. Chang and Q. K. Xue for valuable discussions. This work is supported
by the Defense Advanced Research Projects Agency Microsystems
Technology Office, MesoDynamic Architecture Program (MESO) through the contract number
N66001-11-1-4105, the DARPA Program on ``Topological Insulators -- Solid State Chemistry, New Materials and Properties".
under the award number N66001-12-1-4034 and by the US Department of
Energy, Office of Basic Energy Sciences, Division of Materials Sciences and Engineering, under contract
DE-AC02-76SF00515.

\end{document}